\documentclass[aip, jcp, amsmath, amssymb, citeautoscript, reprint]{revtex4-2}
\usepackage{graphicx, float, xcolor, pgf}
\usepackage{physics, siunitx}
\usepackage[caption=false, justification=centerlast]{subfig}

\usepackage{hyperref}
\hypersetup{
    pdfauthor={Amartya Bose},
    pdftitle={Quantum correlation functions through tensor network path integral},
    colorlinks=true,
}

\allowdisplaybreaks

\begin{document}
\title{Quantum correlation functions through tensor network path integral}
\author{Amartya Bose}
\affiliation{Department of Chemical Sciences, Tata Institute of Fundamental Research, Mumbai 400005, India}
\email{amartya.bose@tifr.res.in}
\begin{abstract}
    Tensor networks have historically proven to be of great utility in providing
    compressed representations of wave functions that can be used for
    calculation of eigenstates. Recently, it has been shown that a variety of
    these networks can be leveraged to make real time non-equilibrium
    simulations of dynamics involving the Feynman-Vernon influence functional
    more efficient. In this work, tensor networks are utilized for calculating
    equilibrium correlation function for open quantum systems using the path
    integral methodology. These correlation functions are of fundamental
    importance in calculations of rates of reactions, simulations of response
    functions and susceptibilities, spectra of systems, etc. The influence of
    the solvent on the quantum system is incorporated through an influence
    functional, whose unconventional structure motivates the design of a new
    optimal matrix product-like operator that can be applied to the so-called
    path amplitude matrix product state. This complex time tensor network path
    integral approach provides an exceptionally efficient representation of the
    path integral enabling simulations for larger systems strongly interacting
    with baths and at lower temperatures out to longer time. The design and
    implementation of this method is discussed along with illustrations from
    rate theory, symmetrized spin correlation functions, dynamical
    susceptibility calculations and quantum thermodynamics.
\end{abstract}
\maketitle
\section{Introduction}
Equilibrium correlation functions provide deep insight into various quantum
processes. They become especially valuable because of the direct link that they
share with experimentally observable quantities through various response
functions and susceptibilities~\cite{weissQuantumDissipativeSystems2012,
    schurichtDynamicalSpinspinCorrelation2009}. Most spectroscopy can be written as
Fourier transforms of corresponding equilibrium correlation
functions~\cite{mukamelPrinciplesNonlinearOptical1995}. Quantum rate theories
have also been formulated in terms of correlation functions involving the
reactive flux operator~\cite{kuboStatisticalMechanicalTheoryIrreversible1957,
    yamamotoQuantumStatisticalMechanical1960, millerQuantumMechanicalTransition1974,
    millerQuantumMechanicalRate1983}.

However, simulating these correlation functions can be quite challenging and
approximations are regularly invoked~\cite{liuRealTimeCorrelation2007,
    liuGreenFunctionsSpin2016}. The computational cost grows prohibitively with the
number of degrees of freedom. Classical trajectory-based approaches are often
used to approximate thermal correlation functions. Notable amongst these
approaches are the family of semiclassical
methods~\cite{wangForwardBackwardInitial2000,
    wangSystematicConvergenceDynamical2001, millerIncludingQuantumEffects2006,
    liuRealTimeCorrelation2007}, centroid molecular
dynamics~\cite{caoFormulationQuantumStatistical1994Equilibrium,
    caoFormulationQuantumStatistical1994Dynamical}, and ring-polymer molecular
dynamics~\cite{craigQuantumStatisticsClassical2004,
    habershonRingpolymerMolecularDynamics2013}. However, these methods are best
applied to systems where the quantum effects are primarily limited to quantum
dispersion and zero-point energy effects. When quantum tunneling becomes
important, a system-solvent decomposition is often very useful in describing the
dynamics. For small system-solvent coupling various perturbation theory methods
are used~\cite{florensDissipativeSpinDynamics2011,
    schadMajoranaRepresentationDissipative2015, schadMajoranaRepresentationSpin2016,
    liuGreenFunctionsSpin2016}. When the coupling is large, approximations like
non-interacting blip
approximation~\cite{dekkerNoninteractingblipApproximationTwolevel1987,
    leggettDynamicsDissipativeTwostate1987, weissQuantumDissipativeSystems2012}
(NIBA) are used. However, these approximations do not provide computationally
feasible routes to systematic improvements of accuracy, and are consequently
limited to particular parameter regimes.

Path integral methods have been derived for open quantum systems that enable the
evaluation of correlation functions particularly in the context of calculating
reaction rates. These methods are numerically exact and not subject to
\textit{ad hoc} approximations. They broadly fall into two categories: (1) Path
integral Monte Carlo-based and analytical continuation-based
methods~\cite{rabaniQuantumMechanicalCanonical2000,simQuantumRateConstants2001}
and (2) quadrature-based.~\cite{topalerQuasiadiabaticPropagatorPath1993,
    topalerQuantumRatesDouble1994, topalerPathIntegralCalculation1996,
    simQuantumRateConstants2001} These quadrature-based methods are typically
derived on top of the quasi-adiabatic propagator path integral (QuAPI) and do
not suffer from the dynamical sign problem that plagues the Monte Carlo
approaches. This has enabled application of the method to calculation of
reaction rates~\cite{topalerQuantumRatesDouble1994,
    shaoIterativePathIntegral2001, shaoIterativePathIntegral2002,
    boseNonequilibriumReactiveFlux2017, boseQuasiclassicalCorrelationFunctions2019}.
However, despite the benefits, owing to the increase in non-Markovian memory
length and a corresponding exponential growth of computational complexity and
memory requirements, QuAPI can only be applied to relatively small systems.

Tensor networks have been applied in a variety of ways to compress and estimate
eigenstates of Hamiltonians through density matrix renormalization group
(DMRG)~\cite{whiteDensityMatrixFormulation1992,
    schollwockDensitymatrixRenormalizationGroup2011,
    schollwockDensitymatrixRenormalizationGroup2011a}. They have also shown
incredible versatility in applications of time propagation of wave
functions~\cite{whiteRealTimeEvolutionUsing2004,
    paeckelTimeevolutionMethodsMatrixproduct2019,
    whiteMinimallyEntangledTypical2009} Inspired by the enormous computational
benefits of such approaches, a number of developments have demonstrated the
utility of tensor networks in alleviating some of the complexity of a QuAPI
calculation.~\cite{strathearnEfficientNonMarkovianQuantum2018,
    jorgensenExploitingCausalTensor2019, boseTensorNetworkRepresentation2021,
    boseMultisiteDecompositionTensor2022, bosePairwiseConnectedTensor2022} In
fact, the multi-site tensor network path
integral~\cite{boseMultisiteDecompositionTensor2022} (MS-TNPI) method in
particular seeks to answer the question to what happens if the framework of
time-dependent DMRG were extended to handle non-Markovian dynamics as seen
in the dynamics of systems interacting with a solvent. These and other
developments~\cite{cerrilloNonMarkovianDynamicalMaps2014,
    makriModularPathIntegral2018, makriSmallMatrixPath2020} have enabled recent
simulations of larger systems with long memory
lengths~\cite{kunduRealTimePathIntegral2020, boseTensorNetworkPath2022,
    boseImpactSolventStatetoState2023, boseImpactSpatialInhomogeneity2023}.

A natural curiosity stems from all of the recent development: can a similar
approach using tensor network be utilized to help with calculations of
correlation functions as well? Would the benefits of tensor network compression
be magnified in these equilibrium calculations? This article seeks to answer
both the previous questions in the affirmative. Here, the path integral
formulation of the equilibrium correlation function of an open quantum system is
expressed in terms of tensor networks. A na\"ive implementation of such an idea
suffers from long-range entanglement in the tensor network structures which can
lead to much greater computational complexity. A procedure has to be developed
that incorporates the structure inherent in these correlation functions and is
able to keep the long-ranged correlations to a minimum.

Only complex-time correlation functions are considered here because these are
expected to benefit the most from the tensor network compression. The
computational benefits of the exponentially decaying imaginary parts is quite
well-known and has been previously used in innovative manners in conjunction
with Monte Carlo to increase the time spans of
simulation~\cite{thirumalaiTimeCorrelationFunctions1984,
    bernePathIntegralMonte1986}. In the same spirit, owing to the damping of the
phases in a complex-time propagator, the complex-time tensor network path
integral method introduced here is expected to have an even greater impact than
that of the real-time non-equilibrium methods that preceded it.

This paper is organized as follows. Section~\ref{sec:method} describes design
and implementation of the tensor network for the complex-time correlation
functions. Detailed discussions of the correlations between the points along the
complex time contour and their impact on the structure of the tensor network is
provided. Numerical examples of the complex-time tensor network path integral
(CT-TNPI) is given in Sec.~\ref{sec:results}. Illustrations of the method are
taken from rate theory, thermodynamics and simulations of symmetrized spin
correlation functions. Finally, some future directions are discussed along with
concluding remarks in Sec.~\ref{sec:conclusion}.

\section{Method}\label{sec:method}
The standard quantum correlation function, which forms the basis of several
observables, between operators $\hat{A}$ and $\hat{B}$, is defined as
\begin{align}
    C_{\hat{A}\hat{B}}(t) & = \frac{1}{Z}\Tr\left(\exp(-\beta \hat{H})\hat{A}(0)\hat{B}(t)\right),\label{eq:standard_corr}
\end{align}
where $\beta=\frac{1}{k_B T}$ is the inverse temperature, $Z = \Tr(\exp(-\beta
        \hat{H}))$ is the partition function and $\hat{B}(t) =
    \exp(i\hat{H}t/\hbar)\hat{B}\exp(-i\hat{H}t/\hbar)$ is the Heisenberg
operator propagated to time $t$. Simulating this correlation function
numerically has often seen to be challenging owing to the unmitigated phases in
the Heisenberg operator $\hat{B}(t)$. The interference of these phases leads to
a so-called ``dynamical sign problem'' for Monte Carlo simulations.

\begin{figure}
    \subfloat[Standard correlation function, Eq.~\ref{eq:standard_corr}]{\includegraphics[scale=0.4]{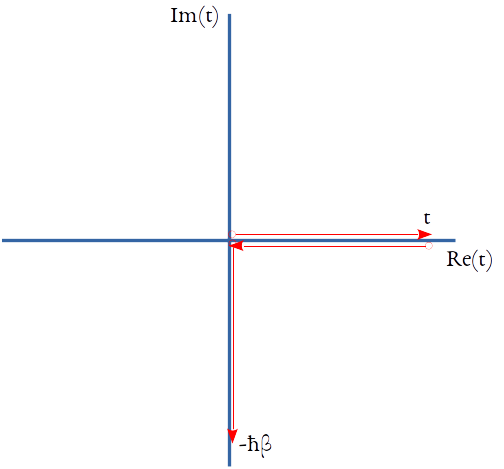}}

    \subfloat[Complex-time correlation function, Eq.~\ref{eq:complex_corr}]{\includegraphics[scale=0.4]{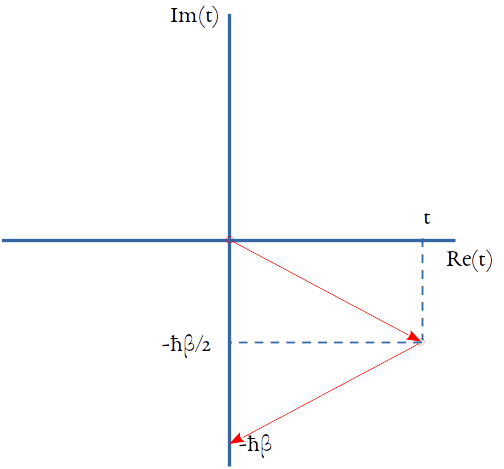}}
    \caption{Time contours for path integral representations of correlation functions.}\label{fig:time_contours}
\end{figure}

To avoid the sign problem, computational focus has primarily been on a related
``complex-time'' correlation function defined by
\begin{align}
    G_{\hat{A}\hat{B}}(t) & = \frac{1}{Z}\Tr\left(\hat{A}(0)\hat{B}(t_c)\right),\label{eq:complex_corr}
\end{align}
where $t_c = t - i\hbar\beta/2$ is the complex time. The phases in
Eq.~\ref{eq:complex_corr} are exponentially dampened by the decaying imaginary
time terms, thereby reducing the dynamical sign problem. This complex-time
correlation function contains the same dynamical information as the standard
correlation function, Eq.~\ref{eq:standard_corr}, and is related to it in the
Fourier domain by
\begin{align}
    G_{\hat{A}\hat{B}}(\omega) & = \exp\left(-\frac{\hbar\beta\omega}{2}\right) C_{\hat{A}\hat{B}}(\omega).
\end{align}
The time contours corresponding to Eq.~\ref{eq:standard_corr} and
Eq.~\ref{eq:complex_corr} are shown in Fig.~\ref{fig:time_contours}.

The goal is to simulate correlation functions for systems interacting with
$N_\text{env}$ uncorrelated thermal environments. The operators, $\hat{A}$ and
$\hat{B}$, whose correlation function we are interested in, act on the system.
\begin{align}
    \hat{H}                  & = \hat{H}_0 + \sum_b^{N_\text{env}}\hat{H}^{(b)}_\text{env}\label{eq:sys_bath}                                                                                          \\
    \hat{H}^{(b)}_\text{env} & = \sum_j \frac{p_{j,b}^2}{2m_{j,b}} + \frac{1}{2}m_{j,b}\omega_{j,b}^2\left( x_{j,b} - \frac{c_{j,b} \hat{s}_b}{m_{j,b}\omega_{j,b}^2} \right)^2. \label{eq:quapi_bath}
\end{align}
The $b$th bath interacts with the system through the operator $\hat{s}_b$. The
frequencies, $\omega_{j,b}$, and couplings, $c_{j,b}$, are characterized by the
spectral density,
\begin{align}
    J_b(\omega) & = \frac{\pi}{2}\sum_j \frac{c_{j,b}^2}{m_{j,b}\omega_{j,b}}\delta(\omega-\omega_{j,b}).
\end{align}
For molecular or atomistic environments, the corresponding harmonic bath can be
obtained through simulations of the energy-gap correlation
function~\cite{makriLinearResponseApproximation1999,
    boseZerocostCorrectionsInfluence2022}.

To obtain the path integral representation for the thermal complex-time correlation
function, Eq.~\ref{eq:complex_corr}, the correlation function is be expressed
as follows
\begin{align}
    G_{\hat{A}\hat{B}}(t)                   & = \frac{1}{Z}\Tr_\text{sys}\left(\hat{A}\mathbb{O}_{\hat{B}(t_c)}\right),               \\
    \text{where } \mathbb{O}_{\hat{B}(t_c)} & = \Tr_\text{bath}\left(\exp(i\hat{H}t_c^*/\hbar)\hat{B}\exp(-i\hat{H}t_c/\hbar)\right).
\end{align}
Once we are able to simulate $\mathbb{O}_{\hat{B}(t_c)}$ for arbitrary
$\hat{B}$, $Z$ can be obtained by setting $\hat{B}$ to the identity operator.

The final path integral expression for $\mathbb{O}_{\hat{B}(t_c)}$ for N
time steps at a time point $t_c = N\Delta t_c$ is given as:
\begin{align}
    \mel{s_1}{\mathbb{O}_{\hat{B}(t_c)}}{s_{2N+2}} & = \sum_{s_2}\sum_{s_3}\cdots\sum_{s_{2N+1}} \mel{s_1}{U^\dag}{s_2}       \nonumber               \\
                                                   & \times\mel{s_2}{U^\dag}{s_3} \cdots\mel{s_N}{U^\dag}{s_{N+1}}            \nonumber               \\
                                                   & \times\mel{s_{N+1}}{\hat{B}}{s_{N+2}}                                    \nonumber               \\
                                                   & \times\mel{s_{N+2}}{U}{s_{N+3}} \cdots\mel{s_{2N}}{U}{s_{2N+1}}          \nonumber               \\
                                                   & \times\mel{s_{2N+1}}{U}{s_{2N+2}}\times F\left[\left\{s_j\right\}\right]\label{eq:path_integral} \\
    = \sum_{s_2}\sum_{s_3}\cdots\sum_{s_{2N+1}}    & P^{(0)}_{s_1,s_2,\ldots,s_{2N+2}} F\left[\left\{s_j\right\}\right]
\end{align}
where $U = \exp(-i\hat{H}\Delta t_c/\hbar)$, $U^\dag = \exp(i\hat{H}\Delta
    t_c^*/\hbar)$ is the short time propagator corresponding to a time step of
$\Delta t_c = (t - i\hbar\beta/2) / N$. These bare complex-time propagators can be
grouped together to form the bare path amplitude tensor,
$P^{(0)}_{s_1,s_2,\ldots,s_{2N+2}}$. The total influence functional, $F$, is a product
of the influence functionals corresponding to the each of the uncorrelated
baths.
\begin{align}
    F\left[\left\{s_j\right\}\right]       & = \prod_{b=1}^{N_\text{env}} F_b\left[\left\{s_j\right\}\right]\label{eq:complex_infl}                 \\
    F^{(b)}\left[\left\{s_j\right\}\right] & = \exp\left(-\frac{1}{\hbar}\sum_{k=1}^{2N+2}\sum_{k'=1}^{k}I^{(b)}_{kk'}s^{(b)}_k s^{(b)}_{k'}\right)
\end{align}
where $s^{(b)}_k$ is the eigenvalue of $\hat{s}_b$ corresponding to $s_k$. The
diagonal terms of the influence functional matrix are
\begin{align}
    I^{(b)}_{kk} & = \frac{2}{\pi}\int_0^\infty \dd\omega \frac{J(\omega)}{\omega^2\sinh(\hbar\beta\omega/2)}\nonumber                                        \\
                 & \times \sin\left(\omega \left(\frac{t_{k+1} - t_k - i\hbar\beta}{2}\right)\right)\sin\left(\omega\left(\frac{t_{k+1}-t_k}{2}\right)\right)
\end{align}
and the off-diagonal terms are
\begin{align}
    I^{(b)}_{kk'} & = \frac{4}{\pi}\int_0^\infty \dd\omega \frac{J(\omega)}{\omega^2\sinh(\hbar\beta\omega/2)}\nonumber                           \\
                  & \times \cos\left(\omega\left(\frac{t_{k+1} + t_k - t_{k'+1} - t_{k'} - i\hbar\beta}{2}\right)\right)\nonumber                 \\
                  & \times \sin\left(\omega \left(\frac{t_{k+1} - t_k}{2}\right)\right)\sin\left(\omega\left(\frac{t_{k+1}-t_k}{2}\right)\right),
\end{align}
when $k\ne k'$~\cite{topalerQuasiadiabaticPropagatorPath1993,
    topalerQuantumRatesDouble1994}. This symmetric $I^{(b)}$-matrix is a
discretization of the complex-time bath response function:
\begin{align}
    \alpha^{(b)}(\tau) & = \sum_j \frac{c_{j,b}^2}{2m_{j,b}\omega_{j,b}}\frac{\cos\left(\omega_j (\tau + i\frac{\hbar\beta}{2})\right)}{\sinh\left(\frac{\hbar\beta\omega_j}{2}\right)} \\
                       & = \frac{1}{\pi}\int_0^\infty\dd\omega J_b(\omega)\frac{\cos\left(\omega (\tau + i\frac{\hbar\beta}{2})\right)}{\sinh\left(\frac{\hbar\beta\omega}{2}\right)}
\end{align}
along the contour shown in Fig.~\ref{fig:time_contours}~(b).  As observed by~
\citet{shaoIterativePathIntegral2001} the response function is maximum in the
neighborhood of $t_c = 0$ and $t_c = -i\hbar\beta$. Unlike the real time bath
correlation function~\cite{makriTensorPropagatorIterativeI1995}, though the
complex-time $\alpha$ is finite everywhere, it does not decay with increasing
$\mid t_c\mid$. However, notice that $\alpha$ decays as $\Re(t_c)\to\infty$ and
as $\Im(t_c)\to -i\hbar\beta/2$. The goal is to use this localization of
non-Markovian interactions to generate compact tensor network representations of
the path amplitude tensor.

\begin{figure}
    \subfloat[Real part of $\alpha$.]{\includegraphics{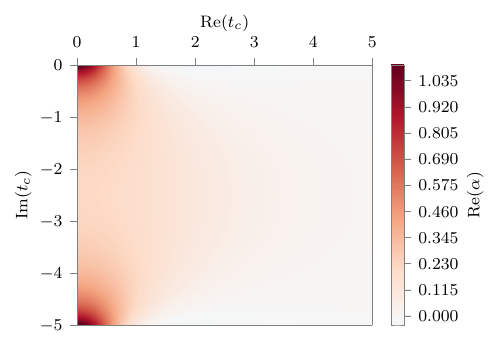}}

    \subfloat[Imaginary part of $\alpha$.]{\includegraphics{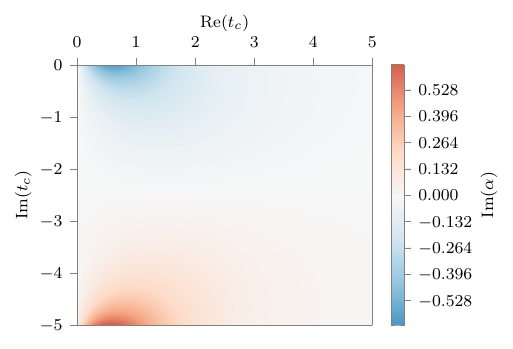}}
    \caption{Bath response function $\alpha$ as a function of complex time for an Ohmic spectral density with $\xi = 2$ and $\omega_c = 1$ at an inverse temperature of $\beta = 5$.}\label{fig:alpha}
\end{figure}

\begin{figure}[t]
    \includegraphics[scale=0.5]{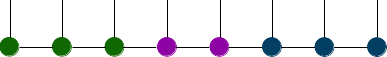}
    \caption{Path amplitude matrix product state. Green sites correspond to those with $U^\dag$, purple to those with $\hat{B}$, and blue with those with $U$.}\label{fig:pa_mps}
\end{figure}

To begin the process observe that the bare path amplitude tensor only connects
the nearest neighbors. Consequently, a matrix product representation of this
tensor should be highly efficient. Consider the singular value factorizations
of $U$, $U^\dag$ and $\hat{B}$:
\begin{align}
    U^\dag_{s_j, s_{j+1}}      & = \sum_{\alpha_j}L^{(\text{back})}_{s_j, \alpha_j} R^{(\text{back})}_{\alpha_j, s_{j+1}} \\
    U_{s_j, s_{j+1}}           & = \sum_{\alpha_j} L^{(\text{for})}_{s_j, \alpha_j} R^{(\text{for})}_{\alpha_j, s_{j+1}}  \\
    \hat{B}_{s_{N+1}, s_{N+2}} & = \sum_{\alpha_{N+1}} B^{(L)}_{s_{N+1}, \alpha_{N+1}} B^{(R)}_{\alpha_{N+1}, s_{N+2}}
\end{align}
In the above expressions, the singular value matrices have been absorbed into
either the left or the right matrix. These factorized forms can be reassembled
to create the matrix product state representation of the bare path amplitude
tensor as follows:
\begin{align}
    P^{(0)}_{s_1,s_2,\ldots,s_{2N+2}}               & = \sum_{\{\alpha_j\}}M^{(1)}_{s_1, \alpha_1} M^{(2)}_{\alpha_1, s_2, \alpha_2} \cdots M^{(2N+2)}_{\alpha_{2N+1}, s_{2N+2}} \\
    \text{where }M^{(1)}_{s_1, \alpha_1}            & = L^{(\text{back})}_{s_1, \alpha_1},                                                                                       \\
    M^{(j)}_{\alpha_{j-1}, s_{j}, \alpha_{j}}       & = R^{(\text{back})}_{\alpha_{j-1}, s_{j}} L^{(\text{back})}_{s_{j}, \alpha_{j}}, 2\le j\le N,                              \\
    M^{(N+1)}_{\alpha_N, s_{N+1}, \alpha_{N+1}}     & = R^{(\text{back})}_{\alpha_N, s_{N+1}} B^{(L)}_{s_{N+1}, \alpha_{N+1}},                                                   \\
    M^{(N+2)}_{\alpha_{N+1}, s_{N+2}, \alpha_{N+2}} & = B^{(R)}_{\alpha_{N+1}, s_{N+2}} L^{(\text{for})}_{s_{N+2}, \alpha_{N+2}},                                                \\
    M^{(j)}_{\alpha_{j-1}, s_{j}, \alpha_{j}}       & = R^{(\text{for})}_{\alpha_{j-1}, s_{j}} L^{(\text{for})}_{s_{j}, \alpha_{j}}, N+3\le j\le 2N+1,                           \\
    M^{(2N+2)}_{\alpha_{2N+1}, s_{2N+2}}            & = R^{(\text{for})}_{\alpha_{2N+1}, s_{2N+2}}.
\end{align}
Here, $\alpha_j$ for $1\le j\le 2N+1$ are the bond indices and their
corresponding dimensions are the so-called ``bond dimensions''. Because of the
nearest neighbor nature of the terms in the bare path amplitude tensor in
Eq.~\ref{eq:path_integral}, the maximum of the bond dimensions will be equal to
the dimensionality of the system. The $\hat{B}$ operator has already been
included in the bare path amplitude MPS at sites $N+1$ and $N+2$. The path
amplitude MPS is schematically shown in Fig.~\ref{fig:pa_mps}.

Now, the complex time influence functional needs to be incorporated. In order to
accomplish this, we refactorize the total influence functional,
Eq.~\ref{eq:complex_infl}, as a constrained product of terms representing
interactions of one specific site, say $k$, with all other sites,
\begin{align}
    F_k\left[\left\{s_j\right\}\right] & = \exp\left(-\frac{1}{\hbar}\sum_{k'=1}^{2N+2}\sum_{b=1}^{N_\text{env}} I^{(b)}_{kk'} s^{(b)}_k s^{(b)}_{k'}\right).\label{eq:site_infl}
\end{align}
In a direct implementation, there will be a double
counting of every interaction in the product of terms like
Eq.~\ref{eq:site_infl}. This double counting of interactions is avoided by
tracing over the $k$th site immediately after applying $F_k$. The ``constrain''
is that the sum over $k'$ in Eq.~\ref{eq:site_infl} will happen over the
existing sites only. As a consequence, now when we move to, say, site $k+1$,
$F_{k+1}$ will not include the interaction between $k$ and $k+1$ because that
has already been incorporated in the previous step. The procedure will be
continued for the next chosen site. To be able to do this, there are two
necessary problems that we have to solve: (1) we need to be able to provide a
matrix product representation for the interaction of all sites with the $k$th
site, and (2) we need to come up with an ordering of the sites that minimizes
the computational and storage requirements.  In this, sites 1 and $2N+2$ will
not be traced over as they are required for the matrix representation of
$\mathbb{O}_{\hat{B}(t_c)}$ and subsequent multiplication with $\hat{A}$.

\begin{figure}[t]
    \includegraphics[scale=0.5]{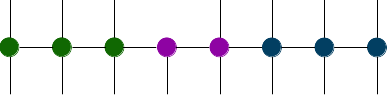}
    \caption{Schematic of the first influence functional operator applied to the path amplitude MPS. The colors are only for helping synchronize this schematic with Fig.~\ref{fig:pa_mps}. Note that the site for which all interactions are included does not have an upper edge indicating an automatic trace over the coordinate.}\label{fig:first_infl}
\end{figure}

Let us start with the second problem: assuming that we can represent the
influence functional operator of all the interactions with site $j$, how do we
order the application of these operators for maximal efficiency? Are there
orderings that are better than others? As discussed, Fig.~\ref{fig:alpha}
demonstrates a decay of correlation as $\Re(t_c)$ increases and $\Im(t_c)\to
    -i\hbar\beta/2$. So, if we started applying the operator from site 2 and tracing
it out, the correlations would initially decrease till $\Im(t_c) =
    -i\hbar\beta/2$ but then increase till the end of the contour where $\Im(t_c) =
    -i\hbar\beta$. These highly non-local interactions, that do not decay
monotonically with distance along the complex time contour, would make the bond
dimensions of the MPS grow very quickly on application of the influence
functional MPO. However, we can try to apply the influence functional operators
in pairs from the middle outwards. This has the advantage of making the
non-local interactions ``look'' short-ranged, and consequently prevents the
spurious growth of bond dimension in any other ordering. In fact, given the
structure of $\alpha$, going from $t_c = -i\hbar\beta/2$ outwards towards $t_c =
    0$ and $t_c=-i\hbar\beta$ is computationally the most optimal order of
incorporation of the influence functional.

We will think in terms of multiple steps of incorporation of the influence
functional, with each step consisting of incorporation of the non-Markovian
interactions involving the middle two points. So, the first operator applied
would take care of all non-Markovian interactions with site $N+1$ and tracing
over it. A schematic of this is shown in Fig.~\ref{fig:first_infl}. This is not
a typical matrix product operator (MPO). A key difference being that MPOs have
the same number of downward and upward site indices. Here, the site whose
interactions are being taken into account does not have an upward index.
Consequently, the output MPS after application of the influence functional
operator has one site less. On application of this operator to the bare path
amplitude MPS will result in an MPS with one less site.

After application of the first influence functional operator, site number $N+1$
ceases to exist. (To keep simplify our nomenclature, we will keep using the
original numbering. So, though site $N+2$ which has the information from the
$\hat{B}$ operator is now the $(N+1)$th site, we will still refer to it as site
$N+2$.  There is no site $N+1$ now.) Then the influence functional corresponding
to site number $N+2$ will be applied, with that site being traced over. (The
influence functional corresponding to the $N+2$ will not contain the
interactions between the $N+2$th site and the $N+1$th site because this has been
incorporated already before tracing over the latter. In this fashion, every
subsequent influence functional will only consist of interactions between the
sites that still exist.) This completes the first step of incorporation of the
non-Markovian influence functional. Sites $N$ and $N+3$ would be incorporated in
the second step, $N-1$ and $N+4$ in the third, so on and so forth till site $2$
and $2N+1$ are incorporated. The first couple of steps of the influence
functional application is schematically shown in Fig.~\ref{fig:first_few_steps}.

\begin{widetext}
    \begin{minipage}{\textwidth}
        \begin{figure}[H]
            \centering
            \includegraphics[scale=0.4]{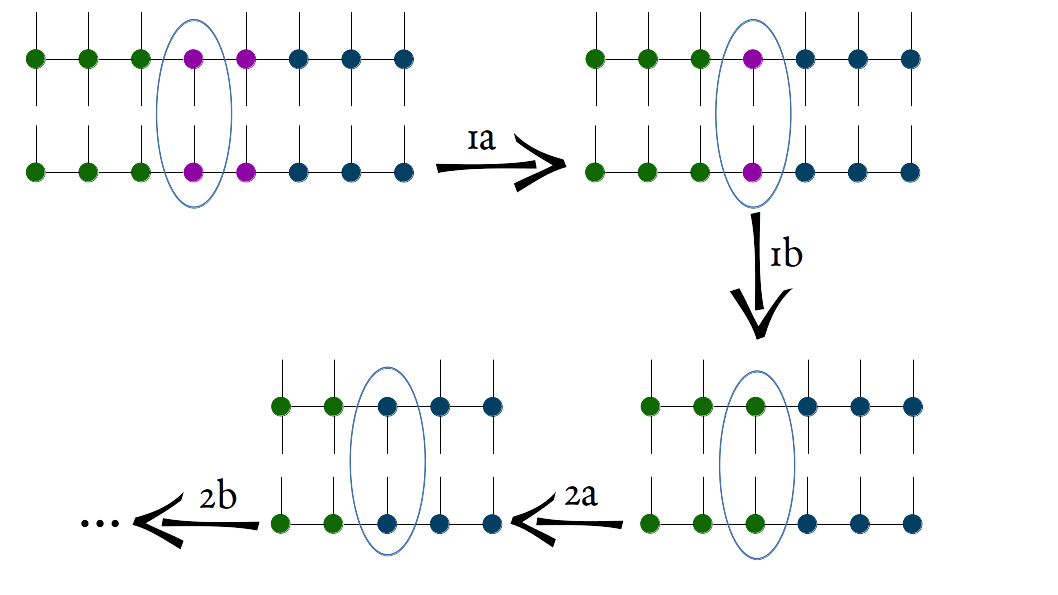}
            \caption{First few steps of application of the influence functional operators. On application of the influence functional operator to the MPS, there is always one site with no dangling edges (no site index). This internal site is absorbed into the site that is left to it.}\label{fig:first_few_steps}
        \end{figure}
    \end{minipage}
\end{widetext}

Having analyzed the structure of the non-Markovian memory and the order of
application of the operators, we now derive the matrix-product representation
for the complex-time influence functional connecting all the sites to a
particular site $k$ (for $2\le k\le 2N+1$). During the steps of application of
the influence functional operators, we trace out sites from the middle of the
contour. Let $l<k$ be the last remaining site to the left of the site $k$, and
$r>k$ be the first site immediately to the right of site $k$. At this stage of
application, all the sites to the left of $l$ ($j\le l$) and right of $r$ ($j\ge
    r$) must exist.
\begin{align}
    F_k\left[\left\{s_j\right\}\right] & = \sum_{\{\beta_j\}}\mathcal{F}^{(1)}_{s_1, s_1', \beta_1} \mathcal{F}^{(2)}_{\beta_1, s_2, s_2', \beta_2} \cdots\nonumber                                             \\
                                       & \times  \mathcal{F}^{(l)}_{\beta_{l-1}, s_l, s_l', \beta_l}\mathcal{F}^{(k)}_{\beta_{l}, s_k, \beta_{r-1}}\mathcal{F}^{(r)}_{\beta_{r-1}, s_r, s_r', \beta_r}\nonumber \\
                                       & \times\cdots \mathcal{F}^{(2N+1)}_{\beta_{2N}, s_{2N+1}, s_{2N+1}', \beta_{2N+1}}\mathcal{F}^{(2N+2)}_{\beta_{2N+1}, s_{2N+2}, s_{2N+2}'}\label{eq:infl_operator}
\end{align}
Notice that the tensor with index $k$ has only one unprimed site index. This is
the feature that enables the automatic tracing over the $k$th site and is shown
in Fig.~\ref{fig:first_infl}. The exact forms of the constituent tensors,
$\mathcal{F}$ are listed in the Appendix~\ref{app:if_tensor}.

Now, we can define the complete algorithm for simulating the correlation
function at a time point $N\Delta t$ is as follows:
\begin{itemize}
    \item Obtain the bare path amplitude MPS corresponding to $\mathbb{O}_{B(t_c)}$ with $2N+2$ sites.
    \item Apply the influence functional operator corresponding to $F_{N+1}$ which encodes all the interactions with the $(N+1)$th time point along the complex time contour.
    \item Apply $F_{N+2}$. This completes the first step of the application of the influence functional and now the path amplitude MPS has $2N$ sites. (We are still following our convention of retaining the time labels of the points.)
    \item Apply $F_{N}$ and $F_{N+3}$. Path amplitude MPS now has $2N-2$ sites.
    \item Repeat till path amplitude MPS has just 2 sites.
    \item Contract this 2-site MPS into a matrix and apply the influence functional between sites 1 and $2N+2$.
    \item Multiply by $\hat{A}$ and trace to get the correlation function.
\end{itemize}

The computational complexity of any MPS-based method is determined by the
maximum bond dimension of the
MPS~\cite{schollwockDensitymatrixRenormalizationGroup2011,
    paeckelTimeevolutionMethodsMatrixproduct2019}. Though the current CT-TNPI
algorithm is not based on conventional MPO-MPS applications, the same
arguments and ideas go through. For every final time point, we have multiple
influence functional operators acting on the path amplitude MPS in sequence
leading to a possible increase in the maximum bond dimension. We define
$\bar{\beta}(t)$ as the average of the maximum bond dimensions encountered over
simulation for a particular time, $t$. This measure governs the computational
requirements of the method. Various common optimizations and singular
value-based filtering schemes that are used in regular MPO-MPS application can
be used here as well. These are typically governed by two convergence
parameters: the cutoff threshold governing the truncated singular value
decomposition and the maximum bond dimension of the resulting MPS. Variational
procedures may also be used for applying the influence functional operator to
the path amplitude MPS.

\section{Results}\label{sec:results}
As numerical illustrations of the method, we provide examples from rate theory,
calculations of spin correlation functions and susceptibilities and some
thermodynamics of the Fenna--Mathhews--Olson complex of photosynthesis. For the
first two sets of examples the system under study can be simply described by a
symmetric spin-half particle or a two-level Hamiltonian:
\begin{align}
    \hat{H}_0 & = -\hbar\Omega\sigma_x,
\end{align}
where $\sigma_j$ for $j$ in $\left\{x, y, z\right\}$ are the spin-1/2 Pauli
matrices. There is one bath that couples to the system through the $\sigma_z$
operator. In these model problems, the spectral density of the bath is taken to
be of the general form:
\begin{align}
    J(\omega) & = \frac{\pi}{2}\hbar\xi\omega^s\omega_c^{1-s}\exp\left(-\frac{\omega}{\omega_c}\right),\label{eq:expcutoff_spectral}
\end{align}
where $\omega_c$ is the characteristic frequency of the bath and $\xi$ is the
dimensionless Kondo parameter encoding the strength of the system-environment
coupling. The parameter $s$ decides the nature of the spectral density, with
$s=1$ being the celebrated Ohmic form.

\subsection{Rate theory}
For the first example consider calculation of reaction rates. Many reactions
happen at significantly slower time scales compared to the ro-translational
motion of the reactants. In such cases, it becomes difficult to directly
simulate the reaction, and one often resorts to calculations of reaction rates.
While reaction rates can be calculated classically, such approaches miss out on
quantum effects of nuclei, and are generally unsuitable for purely non-adiabatic
reactions like electron transport.  It has been shown that the rate of a
reaction is linked to the improper integral of the flux-flux correlation
function over all time, or the zero frequency component of the spectrum
corresponding to the flux-flux correlation
function:~\cite{kuboStatisticalMechanicalTheoryIrreversible1957,
    yamamotoQuantumStatisticalMechanical1960, millerQuantumMechanicalTransition1974,
    millerQuantumMechanicalRate1983}
\begin{align}
    k         & = \int_0^\infty C_{ff}(t)                                                                                                   \\
    C_{ff}(t) & = \frac{1}{Q} \Tr\left(F \exp\left(i\frac{\hat{H}t_c^*}{\hbar}\right) F \exp\left(-i\frac{\hat{H}t_c}{\hbar}\right)\right),
\end{align}
where $Q$ is the reactant partition function and $F=
    \frac{i}{\hbar}\commutator{\hat{H}}{\dyad{R}}$ is the flux operator. An
equivalent formulation can be built in terms of the long-time limit of the
so-called ``flux-side'' correlation, which is the integral of the flux-flux
correlation function:~\cite{millerQuantumMechanicalTransition1974,
    millerQuantumMechanicalRate1983}
\begin{align}
    k         & = \lim_{t\to\infty} C_{fs}(t)                                                                                                      \\
    C_{fs}(t) & = \frac{1}{Q} \Tr\left(F \exp\left(i\frac{\hat{H}t_c^*}{\hbar}\right) \dyad{R} \exp\left(-i\frac{\hat{H}t_c}{\hbar}\right)\right).
\end{align}
In condensed phase reactions, the infinite time limits in both the equations can
be truncated to a plateau time $t_\text{plateau}$ after which the flux-side
correlation function becomes constant and the flux-flux correlation function
becomes zero.

One can think of the rate as half of the zero frequency component of a
frequency-dependent rate function corresponding to the flux-flux autocorrelation
function.~\cite{simQuantumRateConstants2001} These quantum flux-based reaction
rates have been simulated not just using equilibrium correlation functions,
approaches of using non-equilibrium~\cite{boseNonequilibriumReactiveFlux2017}
and near-equilibrium
simulations~\cite{boseQuasiclassicalCorrelationFunctions2019} have also been
shown to give the correct rates.

\begin{widetext}
    \begin{minipage}{\textwidth}
        \begin{figure}[H]
            \subfloat[$s=1$]{\includegraphics{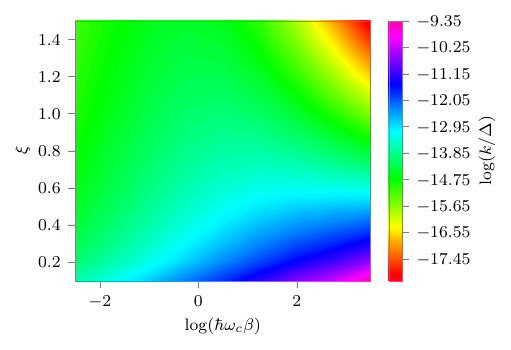}}
            ~\subfloat[$s=0.5$]{\includegraphics{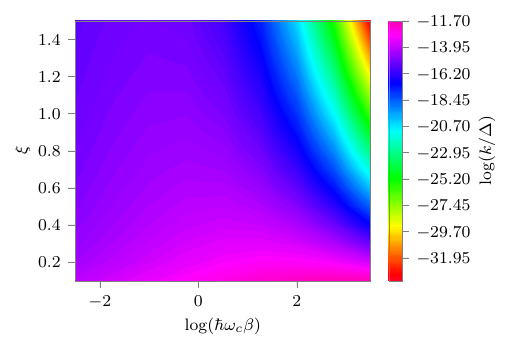}}
            \caption{Rates of reaction for a symmetric two-state system normalized to the tunneling splitting.}\label{fig:reaction_rates}
        \end{figure}
        \begin{figure}[H]
            \subfloat[Flux-side correlation function]{\includegraphics{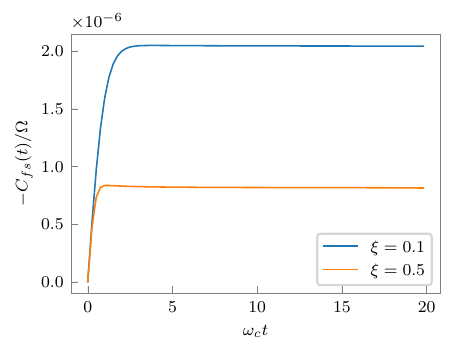}}
            ~\subfloat[Bond dimension as a function of time]{\includegraphics{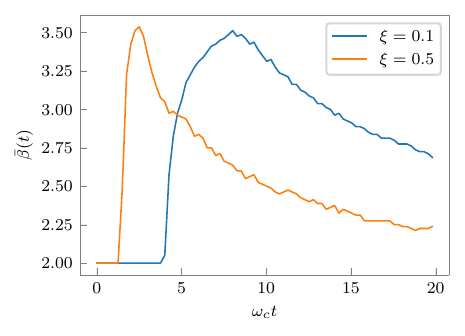}}
            \caption{Simulations for $\hbar\omega_c\beta = 0.2$. Both flux-side correlation function and the values of $\bar{\beta}(t)$ shown for two different system-solvent couplings of $\xi = 0.1$ and $\xi = 0.5$.}\label{fig:performance_flux_side}
        \end{figure}
    \end{minipage}
\end{widetext}

Consider a two-state model for symmetric proton transfer process where the
tunneling splitting is much smaller than the vibronic frequencies. This
separation of time scales, common in many reactions, is critical to the
applicability of rate theory. Following~\citet{topalerQuantumRatesDouble1994},
consider a symmetric two-state system with a tunneling splitting of
$\Delta = 2\hbar\Omega = \SI{0.00105}{\per\cm}$. The environment considered has
the form given in Eq.~\ref{eq:expcutoff_spectral} with a high cutoff frequency
of $\omega_c = \SI{500}{\per\cm}$. We investigate the rates for the Ohmic case
($s=1$) which is ubiquitous for condensed phase systems, a sub-Ohmic case
($s=0.5$), and a super-Ohmic case ($s=3$) which is useful for modeling phonons
in three dimensional solids. The rates for a range of temperatures and coupling
strengths are shown in Fig.~\ref{fig:reaction_rates}. The simulation parameters
used in these cases were much more than what is necessary for convergence. The
data shown in Fig.~\ref{fig:reaction_rates} correspond to $N=30$ and a cutoff
threshold of $10^{-20}$.

To demonstrate the performance of the CT-TNPI method, we plot the flux-side
correlation function and $\bar{\beta}(t)$ for two particular parameters in
Fig.~\ref{fig:performance_flux_side}. These calculations were run with $N=40$
and a cutoff threshold of $10^{-20}$. Despite this, $\bar{\beta}(t)$ is very
small throughout the period of simulation demonstrating the efficiency of the
method. The magnitudes of $\bar{\beta}(t)$ is often dependent on the particular
correlation function as we will numerically demonstrate in the next examples.

\subsection{Symmetrized spin correlation functions}
One of the dynamical quantities of interest for the spin-boson model is the
symmetrized spin correlation function (SSCF)~\cite{liuGreenFunctionsSpin2016,
    schadMajoranaRepresentationSpin2016}. For a generic spin operator, $\sigma$, the
SSCF, defined as
\begin{align}
    S(t) & = \frac{1}{2Z}\Tr\left(\exp(-\beta \hat{H})\left(\sigma(t)\sigma(0) + \sigma(0)\sigma(t)\right)\right),
\end{align}
is related to the structure factor~\cite{weissQuantumDissipativeSystems2012}.
The spectrum of SSCF is related to the spectrum of the complex-time correlation
function as
\begin{align}
    S(\omega) & = \cosh\left(\frac{\hbar\omega\beta}{2}\right) G(\omega),
\end{align}
where $G(\omega) = \int_{-\infty}^\infty G(t)\exp(i\omega t)\dd t$.

Additionally, the Fourier transform of the imaginary part of the dynamical
susceptibility,
\begin{align}
    \chi(t) & = \frac{i\Theta(t)}{Z}\Tr\left(\exp(-\beta \hat{H})\left(\sigma(t)\sigma(0) - \sigma(0)\sigma(t)\right)\right).
\end{align}
is also related to the spectrum of SSCF. This along with Kramer-Kronig's
relation allow us to estimate both the real ($\chi'(\omega)$) and the imaginary
($\chi''(\omega)$) parts of $\chi(\omega)$. Thus, combining all these relations,
it is possible to calculate the symmetrized correlation function as well as the
dynamical susceptibility for open quantum systems directly with the CT-TNPI.

\begin{figure}
    \subfloat[Weakly coupled bath with $\xi = 0.03$.]{\includegraphics{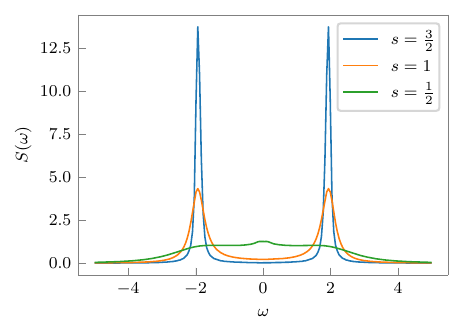}}

    \subfloat[Strongly coupled bath with $\xi = 0.18$.]{\includegraphics{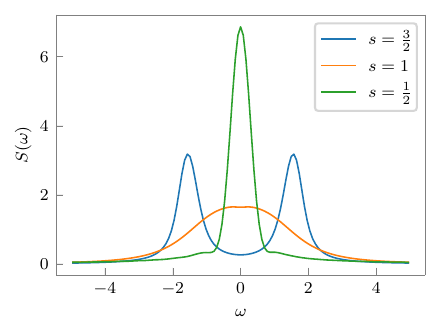}}
    \caption{Comparison of spectra corresponding to SSCF of system coupled with Ohmic and non-Ohmic baths of different strengths.}\label{fig:sscf_spectra}
\end{figure}
\begin{figure}
    \includegraphics{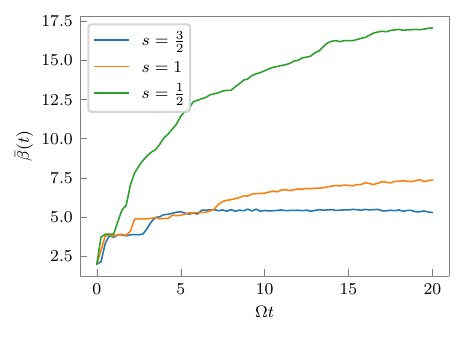}
    \caption{Bond dimension for different baths with Kondo parameter, $\xi = 0.03$.}\label{fig:bond_dim}
\end{figure}
Consider a symmetric spin-boson Hamiltonian with a tunneling splitting
$2\hbar\Omega = 2$ connected to a spectral density given by
Eq.~\ref{eq:expcutoff_spectral} with a high cutoff frequency of $\omega_c =
    30\Omega$ held at a temperature of $\hbar\Omega\beta = 0.4$. We study SSCF
for three different kinds of spectral densities defined by the value of $s$:
we look at the Ohmic spectral density ($s=1$), a sub-Ohmic spectral density
($s=0.5$), and a super-Ohmic spectral density ($s=1.5$). The SSCF spectra are
shown in Fig.~\ref{fig:sscf_spectra} for two different Kondo parameters $\xi$.
The correlation functions were converged for the time span under consideration
with $N=30$ steps and a cutoff threshold of $10^{-10}$. As we go from sub- to
Ohmic to super-Ohmic, the oscillatory character of the dynamics, represented by
the well-separated peaks of the SSCF spectrum, increases. The bond dimensions as
characterized by $\bar{\beta}(t)$ for all the three baths at a Kondo parameter
of $\xi = 0.03$ is shown in Fig.~\ref{fig:bond_dim}. Notice that in addition to
the sub-Ohmic bath being the most efficient at dissipating the oscillatory
dynamics, it also leads to the fastest growth of $\bar{\beta}(t)$.

In Fig.~\ref{fig:susceptibility_ohmic_subohmic}, we track the real and imaginary
parts of the dynamical suscpetibility at different system-solvent coupling
strength for the Ohmic and sub-Ohmic ($s=0.5$) cases. Notice the faster loss of
coherent oscillations in the sub-Ohmic spectral density gets reflected here as
well. The frequency dependencies of the susceptibilities are extremely dependent
on the spectral density of the bath. Because the path integral calculations are
non-perturbative, unlike other approximations~\cite{liuGreenFunctionsSpin2016,
    xuNoncanonicalDistributionNonequilibrium2016}, the current approach gives the
correct result at all parameters upon systematic convergence.
\begin{widetext}
    \begin{minipage}{\textwidth}
        \begin{figure}[H]
            \subfloat[$\chi'(\omega)$ for sub-Ohmic spectral density ($s=0.5$).]{\includegraphics{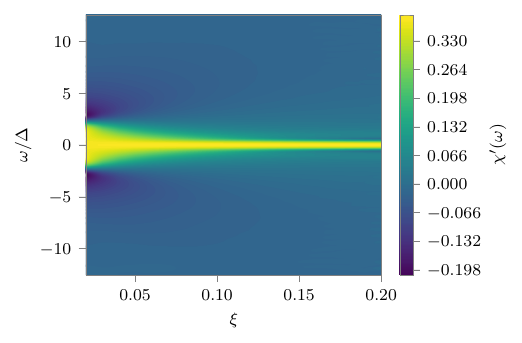}}
            ~\subfloat[$\chi''(\omega)$ for sub-Ohmic spectral density ($s=0.5$).]{\includegraphics{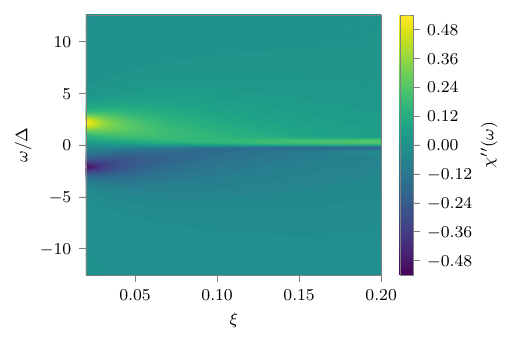}}

            \subfloat[$\chi'(\omega)$ for Ohmic spectral density ($s=1$).]{\includegraphics{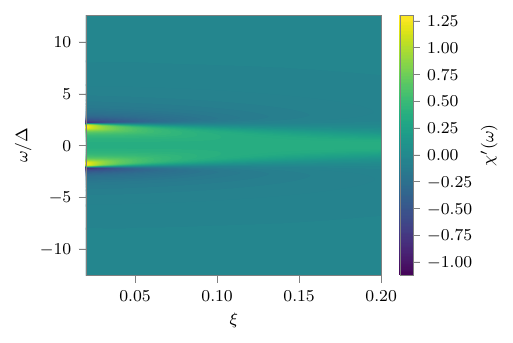}}
            ~\subfloat[$\chi''(\omega)$ for Ohmic spectral density ($s=1$).]{\includegraphics{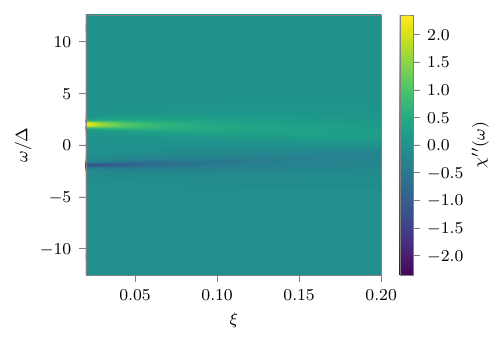}}
            \caption{Dynamical susceptibility for a spin-boson system for different bath spectral densities.}\label{fig:susceptibility_ohmic_subohmic}
        \end{figure}
    \end{minipage}
\end{widetext}

\subsection{Thermodynamics}
As a final example consider the equilibrium density matrices of corresponding to
the single excitation subspace in the Fenna--Matthews--Olson complex (FMO) at an
ambient temperature of $T = \SI{300}{\kelvin}$. FMO is typically a trimer of
octamers of chlorophyll molecules. Here we consider only one unit of the trimer.
The system Hamiltonian is given as
\begin{align}
    \hat{H}_0 & = \sum_j\epsilon_j\dyad{j} + \sum_j\sum_{l\ne j}h_{jl}\left(\dyad{j}{l} + \dyad{l}{j}\right),
\end{align}
where $\epsilon_j$ is the excitation energy of the $j$th chlorophyll molecule,
$h_{jl}$ is the electronic coupling between the excited state of the $j$th
molecule and the ground state of the $l$th molecule. The basis $\ket{j}$
corresponds to the $j$th molecule being in the excited state and all other
molecules in the ground state.

The vibronic couplings happen through localized vibrations and shifts between
the ground and excited Born-Oppenheimer states. The vibrations of the $j$th
molecule have the form:
\begin{align}
    \hat{H}_\text{vib}^{(j)} & =\sum_{i=1}^{N_\text{osc}} \frac{p_{i,j}^2}{2m_{i,j}} + \frac{1}{2}m_{i,j}\omega_{i,j}^2\left( x_{i,j} - \frac{c_{i,j} \dyad{j}}{m_{i,j}\omega_{i,j}^2} \right)^2.
\end{align}
The system Hamiltonian and vibronic degrees of freedom are difficult to
characterize. However recent work~\cite{maityDFTBMMMolecular2020} has provided
some \textit{ab initio} QM/MM level simulations of the spectral densities and
the system Hamiltonians. These descriptions have also been used to study the
excitonic dynamics and pathways of
transport~\cite{boseImpactSolventStatetoState2023,
    boseImpactSpatialInhomogeneity2023}.

\begin{figure}
    \subfloat[TD-LC-DFTB parameterization.]{\includegraphics{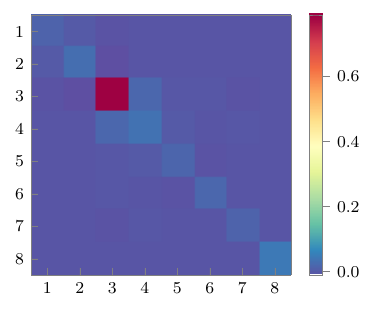}}

    \subfloat[ZINDO parameterization.]{\includegraphics{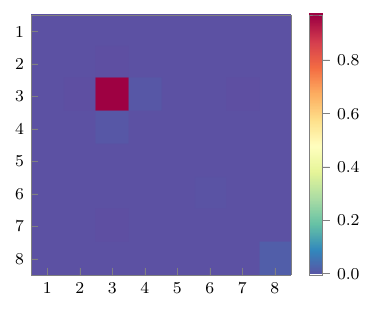}}
    \caption{Equilibrium density matrix for FMO under different parameterizations at $T=\SI{300}{\kelvin}$.}\label{fig:fmo-density-matrix}
\end{figure}

The equilibrium density matrices corresponding to the single excitation sector
is shown in Fig.~\ref{fig:fmo-density-matrix}. Because of differences in the
system Hamiltonian and vibronic couplings, the equilibrium density matrices
shown in Fig.~\ref{fig:fmo-density-matrix} are completely different. It is
interesting that the difference also shows up in the purity of the thermal
ensemble with the ZINDO equilibrium being significantly more pure ($\Tr(\rho^2)
    = 0.938$) than the TD-LC-DFTB equilibrium density matrix ($\Tr(\rho^2)=0.63$).
The von Neumann entropies are also very different with ZINDO having very low
entropy at 0.174 and TD-LC-DFTB being considerably more entangled at 0.91.

\section{Conclusion}\label{sec:conclusion}
Equilibrium correlation functions form the basis for simulations of various
experimentally relevant observables. When open quantum systems are involved,
these correlation functions become to simulate because of an exponential scaling
of computational complexity with number of degrees of freedom. Path integrals
and influence functional provide a lucrative way of rigorously incorporating
influence of baths in these non-Markovian simulations. However as the number of
paths increase, both the computational and storage requirements tend to scale
exponentially.

Tensor network is a commonly-used framework of generating compact
representations of tensors with large orders. Wide-ranging applications of
tensor networks about in physics~\cite{whiteDensityMatrixFormulation1992,
    whiteRealTimeEvolutionUsing2004, whiteMinimallyEntangledTypical2009,
    paeckelTimeevolutionMethodsMatrixproduct2019} and
chemistry~\cite{chanDensityMatrixRenormalization2011,
    renTimeDependentDensityMatrix2018, olivares-amayaAbinitioDensityMatrix2015}.
In this article, we present a tensor network-based approach to simulation of
thermal correlation functions for open quantum systems. A matrix product
state is used to provide a compressed representation for the path ampltiude
tensor. We show that the non-Markovian correlations for the complex-time
influence functional do not decay monotonically with the history. So, if the
influence functional is na\"ively represented as a matrix product operator, the
path amplitude MPS will not offer optimal compression. We use the structure
of the bath response function to motivate an order of applying the influence
functional that minimizes the growth of bond dimension and consequently is
optimal from a computational perspective. Computationally, the tensor network
backend opens up possibilities of future use of graphics processing units to
further accelerate simulations.

Various examples from rate theory, thermodynamics and spin correlation functions
have been used to illustrate the efficiency and broad applicability of the
method. In the near future, we will focus on comparing vibrational dynamics of small
molecules and rates of more complicated reactions in bulk and inside a cavity.
Beyond these system specific studies, this work forms the first step in forming
an infrastructure for multitime correlation functions that leverages tensor
networks for greater efficiency. Such multitime correlation functions are
related to various multidimensional
spectra~\cite{mukamelPrinciplesNonlinearOptical1995}. Furthermore, the
flexibility of the tensor network approaches allow us the ability to explore
routes to equilibration starting from non-equilibrium initial conditions.
Finally, the complex time tensor network path integral code developed in
Julia~\cite{bezansonJuliaFreshApproach2017} using the
\textsc{ITensor}~\cite{fishmanITensorSoftwareLibrary2022,
    fishmanCodebaseReleaseITensor2022} will be released soon as a part of the
QuantumDynamics.jl~\cite{boseQuantumDynamicsJlModular2023} library that was
recently introduced as a platform for simulating dynamics in open quantum
systems using various methods.

\appendix
\section{Influence Functional Operators}\label{app:if_tensor}
Below are the exact forms of the constituent tensors of the matrix product
influence functional operator.
\begin{align}
    \mathcal{F}^{(k)}_{\beta_l, s_k, \beta_{r-1}}         & = \delta_{\beta_l, s_k}\delta_{\beta_{r-1}, s_k} \exp\left(-\frac{1}{\hbar}\sum_{b=1}^{N_\text{env}}I^{(b)}_{kk} s^{(b)}_k s^{(b)}_k\right) \\
    \mathcal{F}^{(k')}_{\beta_{k'-1}, s_{k'}, \beta_{k'}} & = \delta_{\beta_{k'-1}, \beta_{k'}} \exp\left(-\frac{1}{\hbar}\sum_{b=1}^{N_\text{env}}I^{(b)}_{kk'} \beta^{(b)}_{k'} s^{(b)}_{k'}\right)   \\
    \mathcal{F}^{(1)}_{s_{1}, \beta_{1}}                  & = \exp\left(-\frac{1}{\hbar}\sum_{b=1}^{N_\text{env}}I^{(b)}_{1k} \beta^{(b)}_{1} s^{(b)}_{1}\right)                                        \\
    \mathcal{F}^{(2N+2)}_{\beta_{2N+1}, s_{2N+2}}         & = \exp\left(-\frac{1}{\hbar}\sum_{b=1}^{N_\text{env}}I^{(b)}_{k,2N+2} \beta^{(b)}_{2N+1} s^{(b)}_{2N+2}\right)
\end{align}
On application of the influence functional operators, the number of sites on the
path amplitude MPS decreases by one. The extra vertex that is left behind
without a site index is absorbed conventionally in the vertex that is
immediately to the right. With these definitions, we have a complete definition
of the tensor networks required to simulate complex time correlation functions
in a manner that optimally uses the structure of the complex-time bath response
function.

\bibliography{library}
\end{document}